\documentclass[aps,prb,amsmath,amssyb,floatfix,twocolumn]{revtex4}
\usepackage{graphicx}
\usepackage{hyperref}
\usepackage{braket}
\usepackage{bm}

\def\prb{Phys. Rev. B }
\def\prl{Phys. Rev. Lett. }

\def\ie{{\it i.e.}}

\def\etal{{\it et al.}}
\def\bnabla{\bm{\nabla}}
\def\rhos{\rho_s}
\def\D{\mathcal{D}}
\def\na{n_\alpha}
\def\nb{n_\beta}

\def\nt{\tilde{n}}
\def\nta{{\nt_\alpha}}

\def\br{{\bf r}}
\def\bp{{\bf p}}
\def\bq{{\bf q}}
\def\bQ{{\bf Q}}
\def\bA{{\bf A}}
\def\bJ{{\bf J}}
\def\bM{{\bf M}}
\def\ea{{e_\alpha}}

\def\bee{\overline{e^2}}
\def\figscl{0.8}

\begin{document}

\title{Transverse thermoelectric transport
in a model of many competing order parameters}

\author{Gideon~Wachtel and Dror~Orgad}

\affiliation{Racah Institute of Physics, The Hebrew University,
  Jerusalem 91904, Israel}

\date{\today}

\begin{abstract}
  Coexisting fluctuations towards various ordered states are
  ubiquitous in strongly correlated electronic systems. In particular,
  measurements of underdoped cuprate high-temperature superconductors
  reveal evidence for short range charge order in parallel to large
  superconducting fluctuations. Here we use a non-linear sigma model
  to describe a system with $N$ competing orders, and calculate its
  transverse thermoelectric transport coefficient in the analytically
  tractable limit of large $N$.  Our results, which determine the
  contribution of order parameter fluctuations to the Nernst signal,
  are appropriate for high temperatures in the case of finite
  $N$. They are similar to previously obtained results within a model
  of Gaussian superconducting fluctuations.
\end{abstract}

\maketitle

\section{Introduction}

The nature of the psuedogap regime of underdoped cuprate
high-temperature superconductors is still under
debate\cite{Lee}. Studies have conjectured \cite{phase-nature} that
superconducting (SC) fluctuations survive over a large range of
temperatures above the transition temperature, $T_c$. The large Nernst
signal\cite{Huebener,Ongxu,OngPRB01,Onglongprb} measured above $T_c$
has been used to justify this viewpoint, since the Nernst effect is
generally small in non-magnetic metals, and is large in the vortex
state of superconductors. While the Nernst effect in the vortex state
has been well understood for many years\cite{flux-flow}, Ussishkin
\etal\cite{Ussishkin1} were the first to theoretically consider it in
a model of fluctuating superconductivity above $T_c$. They calculated
the transverse thermoelectric transport coefficient within the
Gaussian limit of a time-dependent Ginzburg-Landau (TDGL) model. For
two-dimensional systems, which constitute the focus of our interest,
this model can describe superconducting fluctuations far above the
Berezinskii-Kosterlitz-Thouless transition temperature,
$T_{BKT}$. Their results agree with Nernst measurements in amorphous
Nb$_{0.15}$Si$_{0.85}$ films\cite{Pourret,Michaeli1} and in overdoped,
but not underdoped cuprates\cite{Ussishkin1}, where phase
fluctuations\cite{Podolsky}, specifically thermally excited
vortices\cite{Raghu,Nernst-vortex}, may be required to explain the
Nernst effect closer to $T_{BKT}$.

Over the last few years, X-ray scattering from the pseudogap state of
underdoped cuprates\cite{Ghiringhelli,Chang,Achkar1,Achkar2,
  Comin1,Comin2,daSilva} has revealed charge density wave (CDW) order
whose strength diminishes upon cooling below $T_c$, thereby indicating
a competition between this order and superconductivity. Hence, it is
desirous to reconsider the Nernst effect within a theory which
incorporates the observed competition. A CDW order can affect the
Nernst signal in a couple of ways. One route\cite{Nernst-stripes} is
via the CDW's effect on quasiparticles, which can in turn change the
measured Nernst signal. The second route, which we consider here, is
its competition with SC fluctuations. Above $T_{BKT}$ the CDW
fluctuations are important in determining the properties of thermally
excited SC vortices\cite{nlsm-vortex} and consequently the size of the
Nernst signal. At even higher temperatures, where thermally excited
vortices begin to overlap, Ussishkin \etal's results for the Gaussian
TDGL are expected to hold, provided one properly accounts for the
effect of the competing CDW.

Recently, Hayward \etal\cite{Hayward1} formulated the competition
between the SC and CDW order parameters using a phenomenological
non-linear sigma model (NLSM). By running Monte-Carlo simulations of
their model, they were able to reproduce the temperature dependence of
the CDW structure factor as observed in the X-ray experiments. In
addition, they treated the model analytically in the case of a large
number, $N$, of order parameter components. Using a saddle-point
approximation and including $1/N$ corrections they were able to
reproduce the numerical results for the CDW structure factor and to
calculate the diamagnetic susceptibility at high temperatures. Their
result for the latter agrees with the expected behavior from Gaussian
SC fluctuations.

The transverse thermoelectric transport coefficient, $\alpha_{xy}$, is
defined as the ratio between an applied temperature gradient,
$-\partial_yT$, and the resulting transverse electric current, $J_x$,
\ie, $J_x=\alpha_{xy}(-\partial_yT)$.  For systems with particle-hole
symmetry or when SC fluctuations dominate, the experimentally measured
Nernst signal is given by\cite{Onglongprb} $e_N=\rho\alpha_{xy}$,
where $\rho$ is the longitudinal resistivity.  The purpose of this
paper is to calculate $\alpha_{xy}$ at high temperatures using the
$N\to\infty$ limit of Hayward \etal's model.  For simplicity we
consider the fully $O(N)$ symmetric case, but the results can be
generalized to a more experimentally relevant model, where the
symmetry is not exact. Unlike the magnetization, which is calculated
in equilibrium, the Nernst effect is a transport phenomenon which must
be addressed within a dynamical model. Here we assume that the SC and
CDW fields obey a (Model A) generalized Langevin
equation\cite{HH}. Using a path integral
approach\cite{DeDominicisPeliti} to the Martin-Siggia-Rose
formalism\cite{MSR}, we calculate diagrammatically the system's
response to weak perturbations. As expected, we find that
$\alpha_{xy}$ agrees with Ussishkin's results for the Gaussian TDGL
model. We chose to include here the complete and detailed calculation,
as we believe it has pedagogical value of its own.

The paper is outlined as follows. The model, its Langevin dynamics and
the path integral approach that we use are presented in section
\ref{sec:model}. Section \ref{sec:sp} describes the saddle-point
approximation which is employed throughout the paper. In section
\ref{sec:dia} we summarize various diagrams which are then used to
calculate the diamagnetic susceptibility, in section \ref{sec:Mz}, and
$\alpha_{xy}$ in section \ref{sec:alpha}. We conclude with a
discussion in section \ref{sec:discussion}. Some details of the
calculation are relegated to the appendices.

\section{Model and Dynamics}

\label{sec:model}

We start by considering an $O(N)$ symmetric Ginzburg-Landau model of
$N$ real order parameters, $\na$, $\alpha=1\dots N$, competing with
each other:
\begin{equation}
  \label{eq:GLapp}
  F=\int d^2r\left\{\frac{\rhos}{2}\sum_\alpha(\bnabla\na)^2
    +\frac{u}{4N}\left(\sum_\alpha \na^2-N\right)^2\right\}.
\end{equation}
The non-linear sigma model (NLSM), with the constraint,
\begin{equation}
  \label{eq:constr}
  \sum_\alpha\na^2=N,
\end{equation}
is obtained\cite{PodolskyNLSM} from (\ref{eq:GLapp}) by taking the
limit $u\to\infty$. In order to study transport phenomena we need to
introduce dynamics into the model. A simple approach, which we follow
here, is to assume that the order parameters $\na$ obey stochastic
dynamics, without any conservation constraints (model A of
Ref. \onlinecite{HH}). Thus, the time dependence of the fields $\na$
is given by a generalized Langevin equation,
\begin{equation}
  \label{eq:Lan}
  \frac{\partial\na}{\partial t}=
  -\gamma\frac{\delta F}{\delta\na}+\eta_\alpha,
\end{equation}
where $\gamma$ is a relaxation constant, and $\eta_\alpha$ is a
Gaussian white noise term with $\braket{\eta_a(\br,t)}_\eta=0$ and
\begin{equation}
  \label{eq:eta}
  \braket{\eta_\alpha(\br,t)\eta_\beta(\br',t')}_\eta=2\gamma T
  \delta_{\alpha\beta} \delta(\br-\br')\delta(t-t').
\end{equation}
$\braket{\cdots}_\eta$ denotes an average over all realizations of the
noise term $\eta_\alpha$. The noise correlator in Eq. (\ref{eq:eta})
is determined by the requirement that in the absence of external
perturbations the system relaxes into its equilibrium state as given
by the Gibbs distribution.  We show below that this is the case by
comparing our results to those obtained within an equilibrium
treatment of the same model.

The purpose of the following calculation is, ultimately, to calculate
the response of currents to small perturbations. To this end we
consider the generating functional\cite{DeDominicisPeliti}
\begin{eqnarray}
  \label{eq:ZJ}
  Z[J] & = & \Bigg\langle\int\D n\det M\, \delta\left(
      \frac{\partial\na}{\partial t}
      +\gamma\frac{\delta F}{\delta\na}-\eta_\alpha\right) 
    \nonumber \\   & & \qquad\qquad \qquad
    \times \,e^{\int d^2r\,dt\,\sum_\alpha J_\alpha\na}\Bigg\rangle_\eta,
\end{eqnarray}
from which expectation values of various functions of $\na$ may be
obtained by differentiation with resect to $J$. Here, $\det M$ is a
Jacobian determinant, such that
\begin{equation}
  \int\D n\det M\, \delta\left(\frac{\partial\na}{\partial t}
      +\gamma\frac{\delta F}{\delta\na}-\eta_\alpha\right) = 1,
\end{equation}
where the matrix $M$ itself is given by
\begin{eqnarray}
  \label{eq:M}
  \nonumber
  &&\!\!\!\!\!\!\!\!\!\!\!\!\!\!\!M_{\alpha\beta}(\br,t;\br',t')  =  \\
  &&\!\!\!\!\frac{\delta}{\delta\nb(\br',t')}
  \left[\frac{\partial\na(\br,t)}{\partial t}\
  +\gamma\,\frac{\delta F}{\delta\na(\br,t)}-\eta_\alpha(\br,t)\right].
\end{eqnarray}
In Appendix \ref{sec:Jacob} we evaluate the Jacobian determinant, and
show that\cite{DeDominicisPeliti}
\begin{equation}
  \label{eq:detM}
  \det M = \exp\left[{\frac{\gamma u}{2}\left(1+\frac{2}{N}\right)
      \delta(0)\int d^2r\,dt\,
  \sum_\alpha \na^2}\right].
\end{equation}

The path integral in Eq. (\ref{eq:ZJ}), with the appropriate Jacobian,
is constrained such that for each realization of $\eta_\alpha$, only
the configuration which solves the equations of motion,
Eq. (\ref{eq:Lan}), is included. In order to manage the path integral
over the delta functions, which enforce this constraint, we write them
using auxiliary fields, $\nta$,
\begin{widetext}
  \begin{equation}
    Z[J] = \left\langle\int\D n\D\nt\,e^{\int d^2r\,dt\sum_\alpha
        \left\{i\nta\left(\frac{
              \partial\na}{\partial t}+\gamma\frac{\delta F}{\delta\na}
            -\eta_\alpha\right)+\frac{\gamma u}{2}\left(1+\frac{2}{N}
          \right) \delta(0) \na^2+J_\alpha\na\right\}}
    \right\rangle_\eta.
  \end{equation}
  At this stage, it is simple to preform the average over all
  realization of the noise terms, $\eta_\alpha$, yielding
  \begin{equation}
    Z[J] = \int\D n\D\nt\,e^{\int d^2r\,dt\sum_\alpha
      \left\{i\nta\left(\frac{
            \partial\na}{\partial t}+\gamma\frac{\delta F}{\delta\na}
        \right)-\gamma T\nta^2+ \frac{\gamma u}{2}\left(1+\frac{2}{N}\right)
        \delta(0) \na^2 +J_\alpha\na\right\}}.
  \end{equation}
  By substituting the free energy derivative, and rotating
  $i\nta\to\nta$, we can finally write the generating functional
  \begin{equation}
    Z[J]=\int\D n\D\nt\,e^{-S[\nt,n]+\int d^2r\,dt\,\sum_\alpha J_\alpha\na},
  \end{equation}
  in terms of the action
  \begin{equation}
    \label{eq:Sn4}
    S[\nt,n]=-\int d^2r\,dt\left\{\frac{1}{2}\sum_\alpha
      \begin{array}{cc}
        (\nta & \na) \\ &
      \end{array}\left(
        \begin{array}{cc}
          2\gamma T & L^+ \\ L^- & \gamma u \left(1+\frac{2}{N}\right)
          \delta(0)
        \end{array} \right)\left(
        \begin{array}{c}
          \nta \\ \na
        \end{array}\right) + \frac{\gamma u}{N}\sum_{\alpha\beta}
      \nta\na\nb\nb\right\},
  \end{equation}
  where we have defined $L^\pm=\pm\partial/\partial
  t-\gamma\rhos\nabla^2-\gamma u$.  The action, Eq. (\ref{eq:Sn4}),
  contains a quartic term, and therefore cannot be easily used to
  evaluate response functions. Furthermore, we are interseted in
  results for the NLSM, obtained by taking $u\to\infty$, which rules
  out the possibility of treating the quartic term
  perturbatively. Some progress can be made, though, using a
  saddle-point approximation, as is described in the next section.

  \section{Saddle-point approximation}

  \label{sec:sp}

  The quartic interaction term in Eq. \ref{eq:Sn4} can be decoupled by
  introducing two decoupling fields, $\bar\sigma$ and
  $\bar\lambda$,\cite{decouplenote}
  \begin{equation}
    e^{\int d^2r\,dt\,\frac{\gamma u}{N}\sum_{\alpha\beta}\nta\na\nb\nb}
    =\int\D\bar\sigma\D\bar\lambda\,
    e^{\int d^2r\,dt\,\left\{\bar\sigma\sum_\alpha\nta\na+\bar\lambda
        \sum_\beta\nb\nb-\frac{N}{\gamma u}
        \bar\sigma\bar\lambda\right\}}.
  \end{equation}
  The resulting action, which includes also these decoupling fields,
  is
  \begin{equation}
    S[\nt,n,\bar\sigma,\bar\lambda]=-\int d^2r\,dt\left\{\frac{1}{2}\sum_\alpha
      \begin{array}{cc}
        (\nta & \na) \\ &
      \end{array}\left(
        \begin{array}{cc}
          2\gamma T & L^+ + \bar\sigma \\ L^- + \bar\sigma &
          \gamma u \left(1+\frac{2}{N}\right)\delta(0) +2\bar\lambda
        \end{array} \right)\left(
        \begin{array}{c}
          \nta \\ \na
        \end{array}\right) -\frac{N}{\gamma u}\bar\sigma\bar\lambda\right\}.
  \end{equation}
  It is now possible to integrate over $\nta$ and $\na$, leaving us
  with an action that depends only on $\bar\sigma$ and $\bar\lambda$,
  \begin{equation}
    S[\bar\sigma,\bar\lambda] = \frac{N}{2}{\rm Tr}\ln\left(
      \begin{array}{cc}
        2\gamma T & L^+ + \bar\sigma \\ L^- + \bar\sigma &
        \gamma u \left(1+\frac{2}{N}\right)\delta(0) +2\bar\lambda
      \end{array} \right) +\frac{N}{\gamma u}\int d^2r\,dt\,\bar\sigma\bar\lambda.
  \end{equation}
  In the limit $N\to\infty$ the decoupling fields obtain uniform
  values determined by the saddle-point equations
  \begin{eqnarray}
    \label{eq:SPa}
    \nonumber \\
    \frac{\delta S}{\delta\bar\sigma} & = & \frac{N}{2}\int\frac{d^2p\,d\omega}
    {(2\pi)^3}\frac{L^+(\bp,\omega)+L^-(\bp,\omega)+2\bar\sigma}
    {2\gamma T[\gamma u\delta(0)+2\bar\lambda]-[L^+(\bp,\omega)+\bar\sigma]
      [L^-(\bp,\omega)+\bar\sigma]}-\frac{N\bar\lambda}{\gamma u} \;=\;0 \\
    \nonumber \\  
    \label{eq:SPb}
    \frac{\delta S}{\delta\bar\lambda} & = & \frac{N}{2}\int\frac{d^2p\,d\omega}
    {(2\pi)^3}\frac{4\gamma T}
    {2\gamma T[\gamma u\delta(0)+2\bar\lambda]-[L^+(\bp,\omega)+\bar\sigma]
      [L^-(\bp,\omega)+\bar\sigma]}+\frac{N\bar\sigma}{\gamma u} \;=\;0,
    \\  \nonumber 
  \end{eqnarray}
\end{widetext}
where $L^\pm(\bp,\omega)=\mp i\omega+\gamma\rhos p^2-\gamma u$.  The
first saddle-point equation is solved by\cite{FDTnote}
\begin{equation}
   \label{eq:Spsollam}
   \bar\lambda=-\frac{1}{2}\gamma u\delta(0),
\end{equation}
while the second takes the form
\begin{equation}
  \frac{\bar\sigma}{\gamma u} = \int\frac{d^2p}{(2\pi)^2}\frac{\gamma T}
  {\gamma\rhos p^2+\bar\sigma-\gamma u}.
\end{equation}
Defining an inverse correlation length, $m$, such that
\begin{equation}
   \label{eq:defm}
   \bar\sigma = \gamma\rhos m^2+\gamma u,
\end{equation}
gives in the limit $u\to\infty$,
\begin{equation}
  \label{eq:SPnlsm}
  1 = \int\frac{d^2p}{(2\pi)^2}\frac{T/\rhos}{p^2+m^2},
\end{equation}
that is identical to the saddle-point equation derived within an
equilibrium treatment of the NLSM\cite{Hayward1}. This provides
justification for our choice of the noise correlator,
Eq. (\ref{eq:eta}). Eq. (\ref{eq:SPnlsm}) is solved to give
\begin{equation}
   \label{eq:mofT}
   m^2=\Lambda^2\left[\exp\left(\frac{4\pi\rhos}{T}\right)-1\right]^{-1},
\end{equation}
where $\Lambda$ is an ultra-violet cutoff on the momenta. In the
following we assume that $m<\Lambda$, implying $T\lesssim
4\pi\rhos$. At higher temperatures one needs to put the model,
Eq. (\ref{eq:GLapp}), on a lattice.

At this point, it is convenient to shift the decoupling fields, such
that their saddle-point values vanish in equilibrium, \ie,
$\bar\sigma=\sigma+\gamma\rhos m^2+\gamma u$ and
$\bar\lambda=\lambda-\gamma u\delta(0)/2$. By defining
\begin{equation}
  \label{eq:gpm}
  (g^{\pm})^{-1}=\pm\frac{\partial}{\partial t} +\gamma\rhos
  (-\nabla^2+m^2),
\end{equation}
we can write the action in a form
\begin{widetext}
  \begin{eqnarray}
    \label{eq:Ssp}
    S[\nt,n,\sigma,\lambda] & = & -\int d^2r\,dt\left\{\frac{1}{2}
      \sum_\alpha\begin{array}{cc}
        (\nta & \na) \\ &
      \end{array}\left(
        \begin{array}{cc}
          2\gamma T & (g^+)^{-1} + \sigma \\ (g^-)^{-1} + \sigma &
          2\lambda +2\gamma u\delta(0)/N
        \end{array} \right)\left(
        \begin{array}{c}
          \nta \\ \na
        \end{array}\right) \right. \nonumber \\ & &
    \qquad\qquad\qquad\qquad \left.-\frac{N}{\gamma u}(\sigma
      +\gamma\rhos m^2 + \gamma u)(\lambda-\gamma u\delta(0)/2)
    \right\},
  \end{eqnarray}
  which is most appropriate for handling the limit $N\to\infty$. In
  this limit, and in the absence of perturbing forces, the functional
  integrals over $\sigma$ and $\lambda$ are dominated by their saddle
  point configurations $\sigma = \lambda = 0$.  However, when the
  system is perturbed out of equilibrium, as we consider next, these
  values may change. In addition, fluctuations in $\sigma$ and
  $\lambda$ must be considered when extending the calculation to order
  $\mathcal{O}(1/N)$.

  \section{Diagramatic perturbation theory}

  \label{sec:dia}

  In order to couple some of the fields to an electromagnetic
  potential $\bA$, we construct the following complex fields from
  consecutive pairs of real fields
  \begin{equation}
    \psi_\alpha=\frac{1}{\sqrt{2}}(n_{2\alpha-1}+in_{2\alpha})
    \qquad\qquad \phi_\alpha=\frac{1}{\sqrt{2}}(\nt_{2\alpha-1}+i\nt_{2\alpha}).
  \end{equation}
  After minimal coupling to $\bA$, the free energy,
  Eq. (\ref{eq:GLapp}), becomes
  \begin{figure}[t]
    \centering
    \includegraphics[scale=\figscl]{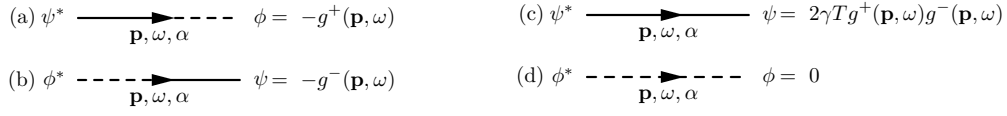}
    \caption{Propagator diagrams. }
    \label{fig:propagators}
  \end{figure}
  \begin{figure}[t]
    \centering
    \includegraphics[scale=\figscl]{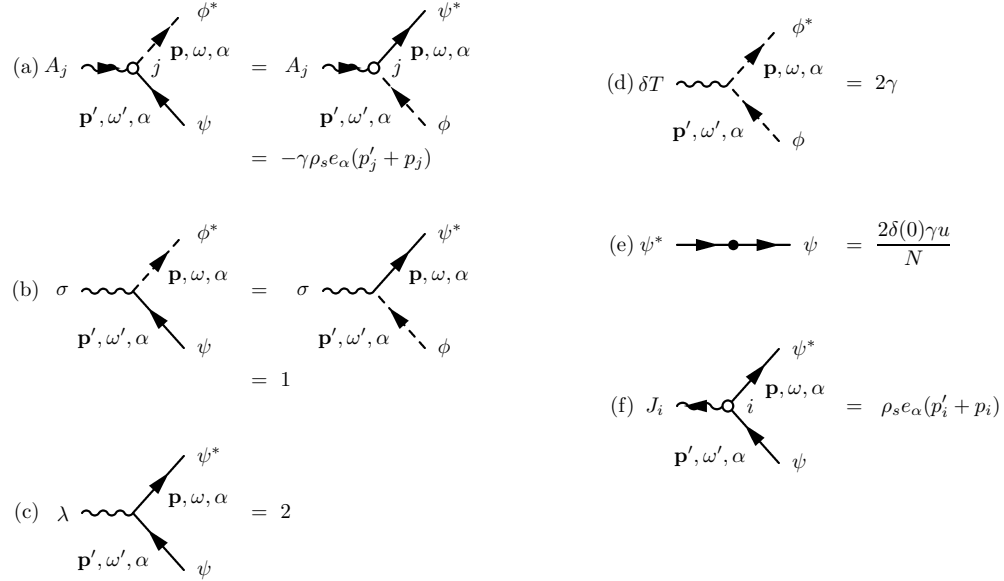}
    \caption{Vertex diagrams. }
    \label{fig:vertices}
  \end{figure}
  \begin{equation}
    \label{eq:GL-psi}
    F=\int d^2r\left\{\rhos\sum_{\alpha=1}^{N/2}\left|(-i\bnabla-\ea\bA)\psi_\alpha
        \right|^2+\frac{u}{4N}\left(2\sum_{\alpha=1}^{N/2}|\psi_\alpha|^2-N\right)^2\right\},
  \end{equation}
  where $\ea$ is the charge of field $\psi_\alpha$, in units where
  $\hbar=c=1$. Similarly, the coupling to $\bA$ and the presence of a
  weak time-dependent temperature gradient $\delta T$ introduce
  additional terms to the action, Eq. (\ref{eq:Ssp}). In preparation
  for constructing a diagramatic perturbation theory, we separate the
  action into two parts $S[\phi,\psi,\sigma,\lambda] =
  S_0[\phi,\psi,\sigma,\lambda] + S_1[\phi,\psi,\sigma,\lambda]$,
  where in the latter we keep only terms linear in $\bA$, hence
  restricting the calculation to linear response
  \begin{eqnarray}
    \label{eq:S0a}
    S_0 & = & -\int d^2r\,dt\sum_{\alpha=1}^{N/2}\left\{
    \begin{array}{cc}
      (\phi^*_\alpha & \psi^*_\alpha) \\ &
    \end{array}\left(
      \begin{array}{cc}
        2\gamma T & (g^+)^{-1}
        \\ (g^-)^{-1}  & 0
      \end{array} \right)\left(
      \begin{array}{c}
        \phi_\alpha \\ \psi_\alpha
      \end{array}\right) -\frac{N}{\gamma u}(\sigma
    +\gamma\rhos m^2 + \gamma u)(\lambda-\gamma u\delta(0)/2)\right\}, \\
    \label{eq:S1}
    S_1 & = & -\int d^2r\,dt\sum_{\alpha=1}^{N/2}
    \begin{array}{cc}
      (\phi^*_\alpha & \psi^*_\alpha) \\ &
    \end{array}\left(
      \begin{array}{cc}
        2\gamma \delta T & \gamma\rhos \ea\{\bA, i\bnabla\}
        + \sigma \\ \gamma\rhos \ea\{\bA, i\bnabla\}
        + \sigma & 2\lambda + 2\gamma u\delta(0) / N
      \end{array} \right)\left(
      \begin{array}{c}
        \phi_\alpha \\ \psi_\alpha
      \end{array}\right). 
  \end{eqnarray}
\end{widetext}

Eq. (\ref{eq:S0a}) defines the $\psi$ and $\phi$ propagators, whose
diagramatic representation is given in Fig. \ref{fig:propagators},
with
\begin{equation}
  \label{eq:g(p,w)psiphi}
  g^{\pm}(\bp,\omega) = \frac{1}{\mp i\omega+\gamma\rho_s(p^2+m^2)}.
\end{equation}
The various interaction terms in Eq. (\ref{eq:S1}) are given by the
vertices in Figs. \ref{fig:vertices}a-e.  To these we add a vertex,
Figure \ref{fig:vertices}f, for the paramagnetic current
$\left.\bJ=-\frac{\delta F}{\delta \bA}\right|_{\bA=0}$, with $F$
given by Eq. (\ref{eq:GL-psi}),
\begin{equation}
  \bJ = i\rhos\sum_{\alpha=1}^{N/2}\ea\left[(\bnabla \psi_\alpha^*)\psi_\alpha -
  \psi_\alpha^*\bnabla\psi_\alpha\right].
\end{equation}

\begin{figure}[b]
  \centering
  \includegraphics[scale=\figscl]{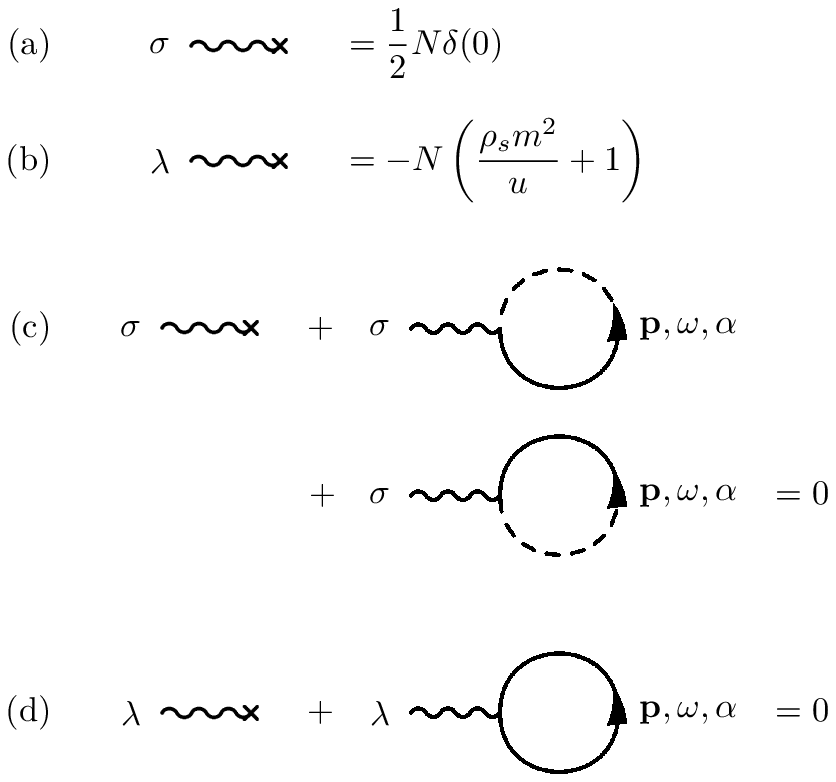}
  \caption{Diagrammatic representation of the saddle-point equations,
    Eqs. (\ref{eq:SPa}) and (\ref{eq:SPb})}.
  \label{fig:saddlepoint}
\end{figure}

\begin{figure}[t]
  \centering
  \includegraphics[scale=\figscl]{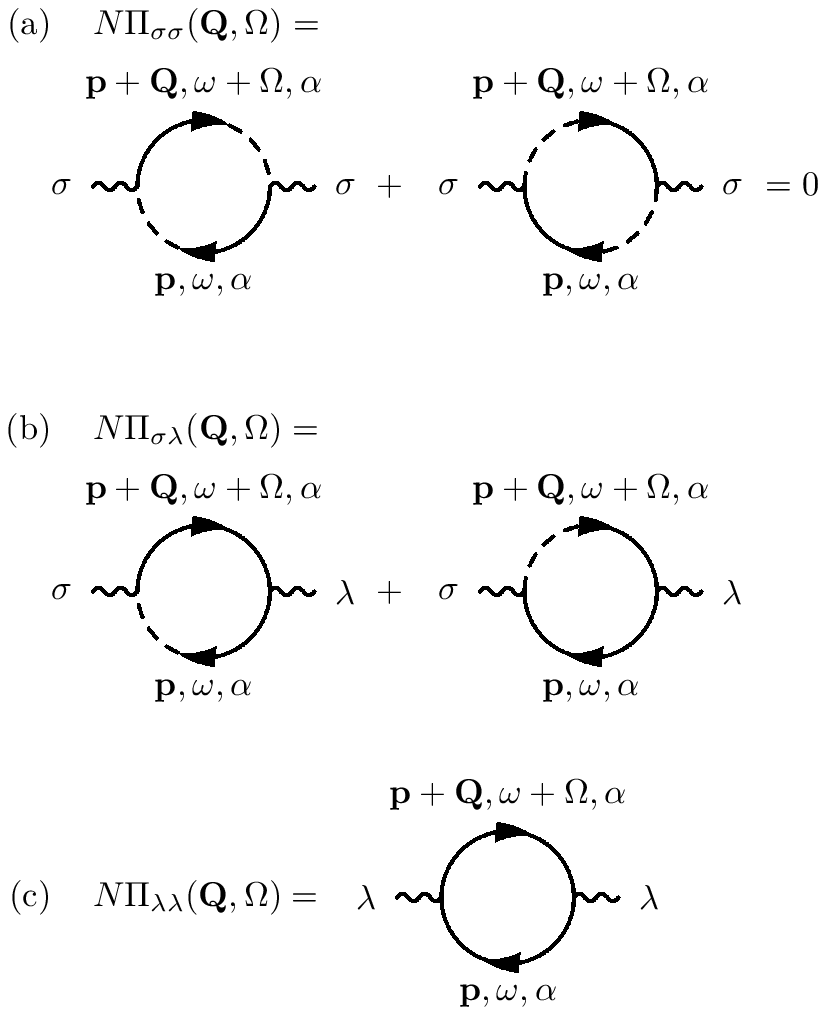}
  \caption{Polarization diagrams for the $\sigma$ and $\lambda$
    fields.}
  \label{fig:bubbles}
\end{figure}

Eq. (\ref{eq:S0a}) also contains source terms for the fields $\sigma$
and $\lambda$, as shown in Fig. \ref{fig:saddlepoint}a-b. However,
they are canceled by the diagrams in Fig. \ref{fig:saddlepoint}c-d, in
what is a diagramatic representation of the saddle-point equations,
Eqs. (\ref{eq:SPa},\ref{eq:SPb}), as can be verified once one notices
that the sum over $\alpha$ runs up to $N/2$ after the model is written
using complex fields. Finally, $S_0$ defines the bare propagators,
$G_0$, for $\sigma$ and $\lambda$. The dressed propagators, to order
$\mathcal{O}(1/N)$, can be constructed using a Dyson equation
$G^{-1}=G_0^{-1}-\Sigma$,
\begin{equation}
  \left(\begin{array}{cc} G_{\sigma\sigma}  & G_{\sigma\lambda}
      \\ G_{\lambda\sigma}  & G_{\lambda\lambda}
    \end{array} \right)^{-1} =
  \left(\begin{array}{cc}  0 & \frac{\gamma u}{N}
      \\ \frac{\gamma u}{N} & 0
    \end{array} \right)^{-1} - N
  \left(\begin{array}{cc} \Pi_{\sigma\sigma}  & \Pi_{\sigma\lambda}
      \\ \Pi_{\lambda\sigma}  & \Pi_{\lambda\lambda}
    \end{array} \right),
\end{equation}
where the polarization diagrams, $\Pi(\bQ,\Omega)$, are given in
Figure \ref{fig:bubbles}.  Since the poles of the bubbles in
Fig. \ref{fig:bubbles}a reside on the same half of the complex
$(\bp,\omega)$ plane we find that $\Pi_{\sigma\sigma}=0$. This leads
in the limit $u\to\infty$ to
\begin{eqnarray}
\label{eq:G}
  \left(\begin{array}{cc} G_{\sigma\sigma}  & G_{\sigma\lambda}
      \\ G_{\lambda\sigma}  & G_{\lambda\lambda}
    \end{array} \right) & = & -\frac{1}{N}
  \left(\begin{array}{cc} 0  & \Pi_{\sigma\lambda}
      \\ \Pi_{\lambda\sigma}  & \Pi_{\lambda\lambda}
    \end{array} \right)^{-1} \nonumber \\ & = & -\frac{1}{N}
  \left(\begin{array}{cc} -\frac{\Pi_{\lambda\lambda}}{(\Pi_{\sigma\lambda})^2}
      & \frac{1}{\Pi_{\sigma\lambda}} \\ \frac{1}{\Pi_{\lambda\sigma}} & 0
    \end{array} \right).
\end{eqnarray}

\section{Diamagnetic susceptibility}

\label{sec:Mz}

\begin{figure}[t]
  \centering
  \includegraphics[scale=\figscl]{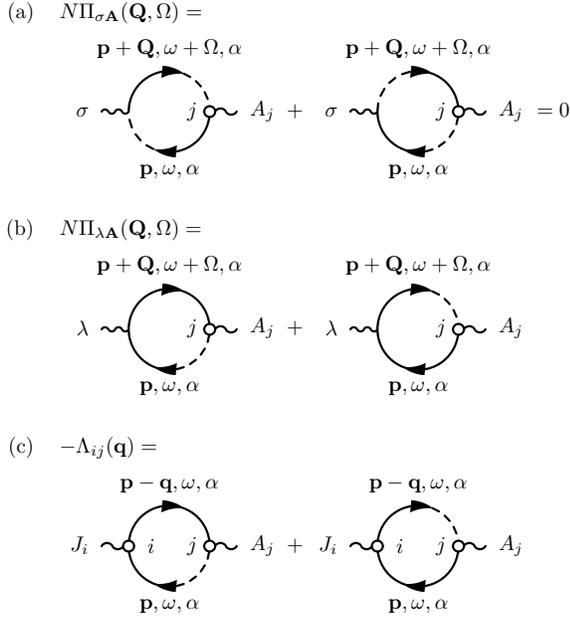}
  \caption{Response of $\sigma$, $\lambda$ and $J_i$ to $A_j$ }
  \label{fig:bubblesA}
\end{figure}

Before calculating the electric current's response to a static weak
magnetic field, we first show that in the presence of such a
perturbation $\sigma$ and $\lambda$ remain unchanged. To this end we
need to calculate the diagrams in Fig. \ref{fig:bubblesA}a-b, which
contain the leading order contribution to the response of $\sigma$ and
$\lambda$ to $A_j$. An examination of the pole structure of
Fig. \ref{fig:bubblesA}a leads to $\Pi_{\sigma\bA}(\bQ,\Omega=0)=0$.
In Appendix \ref{sec:evdia} we explicitly calculate the diagrams
appearing in Fig. \ref{fig:bubblesA}b and find that also
$\Pi_{\lambda\bA}(\bQ,\Omega=0)=0$.

To calculate the magnetization we note that the equilibrium
magnetization currents are given by $\bJ=\bnabla\times\bM$ (this is a
consequence of $\bnabla\cdot\bJ =0$ for the magnetization currents and
is taken as the definition of $\bM$). Therefore, in the $xy$ plane,
$J_i=\varepsilon_{ij}\partial_jM_z$, where $\varepsilon_{ij}$ is the
antisymmetric tensor with $\varepsilon_{xy}=-\varepsilon_{yx}=1$. The
susceptibility $\chi$ is the ratio between $M_z$ and
$B_z=\varepsilon_{ij}\partial_iA_j$. As a result,
\begin{eqnarray}
  \label{eq:chi}
  J_i(\bq) 
  & = &  \chi(\delta_{ij}q^2-q_iq_j)A_j(\bq),
\end{eqnarray}
from which it follows that $\chi$ itself can be calculated by
identifying a term proportional to $q_iq_j$ in the response function
$\Lambda_{ij}(\bq)$, defined via
$J_i(\bq)=-\Lambda_{ij}(\bq)A_j(\bq)$. The diagramatic representation
of the latter is given in Fig. \ref{fig:bubblesA}c and evaluated in
Appendix \ref{sec:evdia}. The result is
\begin{eqnarray}
  \label{eq:Lambda}
  -\Lambda_{ij}(\bq) & = & \frac{\bee T}{2\pi}
  \left[\log\left(\frac{\Lambda^2}{m^2}\right)-1\right]\delta_{ij}
  \nonumber \\
  & & - \frac{\bee T}{12\pi m^2}(\delta_{ij}q^2-q_iq_j),
\end{eqnarray}
where
\begin{equation}
  \bee\equiv\sum_{\alpha=1}^{N/2} e^2_\alpha.
\end{equation}
The $q=0$ piece in Eq. (\ref{eq:Lambda}) should get canceled by the
diamagnetic contribution to the current, which is given by
$-2\bee\rho_s\braket{\psi_\alpha^*\psi_\alpha}\bA$. Its contribution
to $-\Lambda_{ij}$ is
\begin{eqnarray}
  \nonumber
  -2\bee \rho_s\delta_{ij}\int_{\bp\omega}2\gamma Tg^+g^-(\bp,\omega)
  &=&-2\bee  T\delta_{ij}\int_{\bp}\frac{1}{p^2+m^2} \nonumber \\ &=&
  -\frac{\bee  T}{2\pi}\log\left(\frac{\Lambda^2}{m^2}\right)\delta_{ij}.
  \nonumber \\
\end{eqnarray}
The imperfect cancelation is due to the non gauge-invariant cutoff
scheme which we used\cite{Hayward1}. From Eq. (\ref{eq:Lambda}) we
nevertheless obtain
\begin{equation}
  \chi = -\frac{\bee T}{12\pi m^2},
\end{equation}
which is identical to the result calculated using equilibrium methods
in Ref. \onlinecite{Hayward1}. In terms of the inverse correlation
length, $m$, this is also what one finds using the Gaussian
approximation of the Ginzburg-Landau model.

\section{The coefficient $\alpha_{xy}$}

\label{sec:alpha}

The transverse thermoelectric transport coefficient, $\alpha_{xy}$, is
defined via $J_x=\alpha_{xy}(-\partial_yT)$, which we rewrite as
\begin{equation}
   J_i=\frac{\alpha_{xy}}{B}\varepsilon_{lj}\partial_lA_j
  \varepsilon_{ik}(-\partial_kT).
\end{equation}
This Fourier transforms into
\begin{equation}
  J_i(\bq+\bQ) = \frac{\alpha_{xy}}{B}(-\delta_{ij}\bq\cdot\bQ+Q_iq_j)
  T(\bq)A_j(\bQ),
\end{equation}
from which we conclude that the coefficient $\alpha_{xy}/B$ can be
obtained by calculating the response of $J_i$ to $\delta T$ and $A_j$,
and reading off the term proportional to $Q_iq_j$.

\begin{figure}[t]
  \centering
  \includegraphics[scale=\figscl]{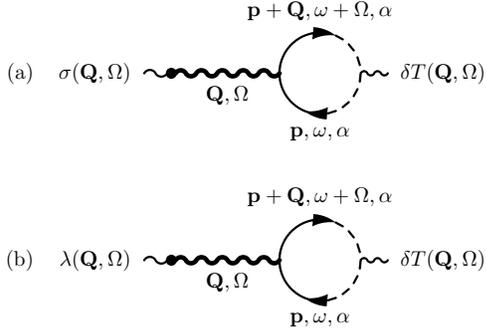}
  \caption{Response of $\sigma$ and $\lambda$ to $\delta T$. The bold
    wavy line represents, $G$, the dressed propagator of $\sigma$ and
    $\lambda$, given by Eq. (\ref{eq:G}).}
  \label{fig:propbub}
\end{figure}

In order to calculate this response, we first examine the change in
the saddle-point values of $\sigma$ and $\lambda$ in the presence of a
slow temperature gradient. The change in $\sigma$ is given by
Fig. \ref{fig:propbub}a,
\begin{eqnarray}
\label{eq:sigmadeltaT}
  \frac{\sigma(\bQ,\Omega)}{\delta T(\bQ,\Omega)}
  &\!\!\! =\!\!\! &
  -\frac{2\gamma}{\Pi_{\sigma\lambda}(\bQ,\Omega)}
  \int_{\bp\omega}\!\!g^-(\bp,\omega)
  g^+(\bp+\bQ,\omega+\Omega), \nonumber \\
\end{eqnarray}
where we have used
$G_{\sigma\lambda}(\bQ,\Omega)=-1/N\Pi_{\sigma\lambda}(\bQ,\Omega)$,
see Eq. (\ref{eq:G}). On the other hand, the temperature derivative of
the saddle-point equation (\ref{eq:SPnlsm}) can be represented as
\begin{eqnarray}
\label{eq:difTSPE}
  0 & = & \frac{d}{dT}\int_{\bp\omega}2\gamma Tg^+(\bp,\omega)g^-(\bp,\omega) \nonumber \\
   & = & \int_{\bp\omega}2\gamma g^+(\bp,\omega)g^-(\bp,\omega) \nonumber \\
  && + 2\gamma T \frac{dm^2}{dT}\frac{d}{dm^2}\int_{\bp\omega}g^+(\bp,\omega)
  g^-(\bp,\omega)\nonumber \\
   & = & \int_{\bp\omega}g^+(\bp,\omega)g^-(\bp,\omega)
   +\frac{\rhos}{2}\frac{dm^2}{dT} \Pi_{\sigma\lambda}(\bQ=0,\Omega=0), \nonumber  \\
\end{eqnarray}
where we have used the relation
\begin{equation}
  \Pi_{\sigma\lambda}(\bQ=0,\Omega=0)=\frac{2T}{\rhos}\frac{d}{dm^2}\int_{\bp\omega}
  g^+(\bp,\omega)g^-(\bp,\omega),
\end{equation}
which is readily verified from the algebraic expression of the
diagrams for $\Pi_{\sigma\lambda}$, Fig. \ref{fig:bubbles}b. Combining
Eqs. (\ref{eq:sigmadeltaT}) and (\ref{eq:difTSPE}) establishes that
{\it in the limit} $\bQ,\Omega\to 0$
\begin{equation}
  \sigma=\gamma\rhos\frac{dm^2}{dT}\delta T.
\end{equation}
Finally, one finds, with the help of Fig.  \ref{fig:propbub}b and
$G_{\lambda\lambda}=0$, that the equilibrium result $\lambda=0$ is
unaffected by the temperature gradient.

\begin{figure}[t]
  \centering
  \includegraphics[scale=\figscl]{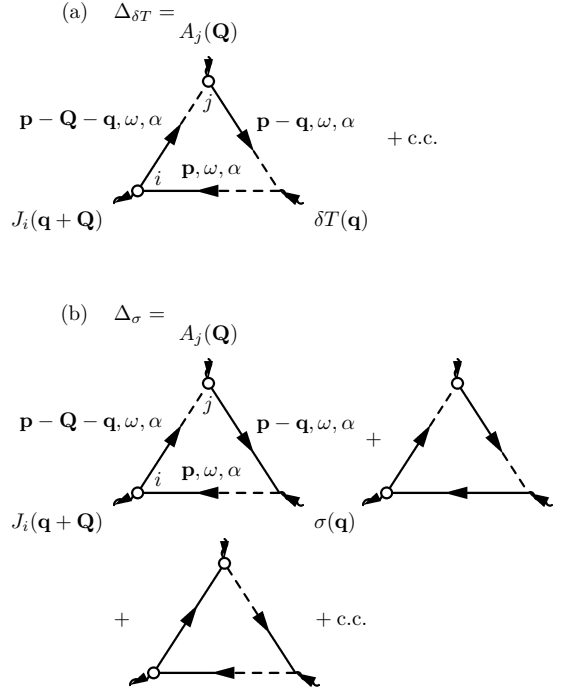}
  \caption{Themoelectric response diagrams.}
  \label{fig:triangles}
\end{figure}

As a consequence of the above discussion, $\alpha_{xy}$ acquires
contributions both from the direct response to a gradient in $\delta
T$ and from the induced change in $\sigma$. The diagrams in
Fig. \ref{fig:triangles}a give the response to $\bnabla\delta T$,
while assuming that $\sigma$ remains at its equilibrium value
$\sigma=0$.  They are calculated in Appendix \ref{sec:evdia} up to
leading order in $Q_iq_j$, with the result
\begin{equation}
  \Delta_{\delta T} \approx
  \delta_{ij}(\cdots)+q_iQ_j(\cdots) + Q_iq_j\frac{\bee }{8\pi m^2}.
\end{equation}
To this we need to add the response to $\bnabla\sigma$, as represented
by the diagrams in Fig. \ref{fig:triangles}b. They are also calculated
in Appendix \ref{sec:evdia}, and up to leading order in $Q_iq_j$ give
\begin{equation}
  \Delta_\sigma \approx  \delta_{ij}(\cdots)+q_iQ_j(\cdots)-Q_iq_j\frac{\bee T}
  {12\pi\gamma\rhos m^4}.
\end{equation}

Part of the response to a temperature gradient in the bulk, $\delta
J_i=\epsilon_{ij}B\frac{d\chi}{dT}\partial_j\delta T$, is due to
redistribution of equilibrium magnetization currents, and should be
subtracted from the calculated bulk current, since it is canceled by
opposite currents on the system edges\cite{Cooper}.  This amounts to
adding $\partial\chi/\partial T$ to the above calculated
$\alpha_{xy}/B$, with the resulting transport response,
\begin{eqnarray}
  \frac{\alpha_{xy}^{\rm tr}}{B} & = & \frac{\partial\Delta_{\delta T}}
  {\partial(Q_xq_y)}+\frac{\partial\Delta_{\sigma}}{\partial(Q_xq_y)}
  \frac{d\sigma}{d\delta T}
  +\frac{\partial\chi}{\partial T} \nonumber \\ & = &
  \frac{\bee }{24\pi m^2} 
  = -\frac{\chi}{2T}.
  \label{eq:alpha}
\end{eqnarray}
This is our main result, which in terms of the inverse correlation
length, $m$, agrees with Ussishkin \etal's result for the Gaussian
TDGL model\cite{Ussishkin1}.

\section{Discussion}

\label{sec:discussion}

While we derived our main result, Eq. (\ref{eq:alpha}), for the fully
$O(N)$ symmetric NLSM, one expects the Hamiltonian of generic systems,
including the underdoped cuprates, to contain terms which explicitly
break the symmetry. It is a simple task to adapt our calculation to a
case where the symmetry breaking terms are quadratic in the
fields. For example, the model may include different values of the
stiffness for different fields, or additional mass terms. In such a
case one only needs to incorporate these changes into the propagators,
via Eq. (\ref{eq:g(p,w)psiphi}). The final result is the same, when
written in terms of the inverse correlation length $m$, which itself
is still a solution of a saddle-point equation, similar to
Eq. (\ref{eq:SPnlsm}), but adapted to the non-symmetric model.  If, on
the other hand, the symmetry breaking terms are of higher order, such
as quartic terms which impose a square lattice point group symmetry on
the CDW components\cite{Hayward1}, then additional decoupling fields
are needed, and the adaptation is not as
straightforward. Nevertheless, it is still reasonable to expect that
the Nernst coefficient's dependence on the inverse correlation length
$m$ remains unchanged.

The saddle point approximation, which we use to arrive at
Eq. (\ref{eq:alpha}), is strictly correct only in the $N\to\infty$
limit. However, it is still expected to describe the model's behavior
at high enough temperatures, since corrections of order $1/N$ become
less important as the temperature is increased\cite{Exnote}.
On the other hand, the approximation is likely to fail at low
temperatures. This is especially true in the physically relevant case
where the symmetry is broken so as to favor SC order. At low
temperatures the symmetry reduces to $O(2)$, and the Nernst signal
should be calculated within a SC vortex based
model\cite{Nernst-vortex}. Thus, as the temperatures is increased, the
Nernst effect is expected to crossover\cite{nlsm-vortex} from vortex
physics at low temperatures, to the Gaussian fluctuations result,
which we calculated here, at high temperatures.

\acknowledgements

This research was supported by the Israel Science Foundation (Grant
No. 585/13).

\appendix

\section{Evaluation of the Jacobian determinant}

\label{sec:Jacob}

Evaluating the Jacobian determinant in the path integral,
Eq. (\ref{eq:ZJ}), can be made simple by separating the matrix
$M_{\alpha\beta}(\br,t;\br',t')$, Eq. (\ref{eq:M}), into two
parts. Defining
\begin{equation}
  \label{eq:P}
  P_{\alpha\beta}(\br,t;\br',t')\equiv
  \frac{\partial}{\partial t}\delta_{\alpha\beta}\delta(\br-\br')\delta(t-t')
\end{equation}
and
\begin{equation}
  \label{eq:Q}
  Q_{\alpha\beta}(\br,t;\br',t')\equiv
  \gamma\frac{\delta^2F}{\delta\na(\br,t)\delta\nb(\br',t')},
\end{equation}
we have
\begin{equation}
  \label{eq:detMap}
  \det M = \det P\,\det(1+P^{-1}Q)=\det P\,e^{{\rm Tr}\ln(1+P^{-1}Q)},
\end{equation}
where
\begin{equation}
  \label{eq:Pinv}
  P^{-1}_{\alpha\beta}(\br,t;\br',t')=\theta(t-t')\delta_{\alpha\beta}\delta(\br-\br').
\end{equation}
$\det P$ is independent of the fields and, as such, can be
disregarded. To evaluate the Jacobian we need only to expand the
logarithm in Eq. (\ref{eq:detMap}) in powers of $P^{-1}Q$.
We find that aside from an irrelevant constant the linear term is given by
\begin{equation}
  {\rm Tr} P^{-1}Q =
  \gamma u\left(1+\frac{2}{N}\right)\theta(0)\delta(0)
  \int d^2r\,dt\,\sum_\alpha\na^2,
\end{equation}
while the quadratic and higher terms vanish\cite{DeDominicisPeliti}.
A proper limiting process gives
\begin{equation}
  \theta(0)=\frac{1}{2} \qquad {\rm and} \qquad
  \delta(0)=\int\frac{d^2p}{(2\pi)^2}.
\end{equation}
where the integral is over the first Brillouin zone.  This establishes
Eq. (\ref{eq:detM}), up to an unimportant normalization constant.
\\

\section{Evaluation of diagrams}

\label{sec:evdia}

In this appendix we calculate in detail those diagrams which are used
in the main text. We begin by showing that $\lambda$ remains zero in
the presence of a static, weak magnetic field. To do so we evaluate
the diagram in Fig. \ref{fig:bubblesA}b for $\Omega=0$,
\begin{widetext}
  \begin{equation}
    N\Pi_{\lambda\bA}(\bQ,\Omega=0) =  4\sum_{\alpha=1}^{N/2} e_\alpha
    \gamma^2\rhos T \int_{\bp\omega}
    (2p_j+Q_j)\left[g^-(\bp,\omega)g^+g^-(\bp+\bQ,\omega)
      + g^+g^-(\bp,\omega)g^+(\bp+\bQ,\omega)\right],
  \end{equation}
  where here and throughout
  $\int_{\bp\omega}\equiv\int\frac{d^2p}{(2\pi)^2}\int\frac{d\omega}{2\pi}$
  etc.  The integrand has only simple poles, so integrating over
  $\omega$ gives
    \begin{eqnarray}
      N\Pi_{\lambda\bA}(\bQ,\Omega=0)
      & = & \frac{2 T}{\rhos} \sum_\alpha e_\alpha  \int_{\bp}\frac{2p_j+Q_j}
      {(p^2+m^2)\left[(\bp+\bQ)^2+m^2\right]} \nonumber \\
      & = &  \frac{2 T}{\rhos}\sum_\alpha e_\alpha\int_\bp\int_0^1du
      \frac{2p_j+Q_j}{\left\{u(p^2+m^2)+(1-u)\left[(\bp+\bQ)^2+m^2\right]
        \right\}^2} \nonumber \\
      & = & \frac{2 T}{\rhos}\sum_\alpha e_\alpha \int_0^1du \int_{\bp'}
      \frac{2p'_j-(1-2u)Q_j}{\left[p'^2+m^2+u(1-u)Q^2\right]^2}=0,
  \end{eqnarray}
  where we have transformed to $\bp'=\bp+(1-u)\bQ$.

  Eq. (\ref{eq:chi}) implies that the response function $\Lambda_{ij}$
  can be used to calculate the diamagnetic susceptibility $\chi$. The
  diagrams for $\Lambda_{ij}(\bq)$ are given in
  Fig. \ref{fig:bubblesA}c, and evaluate to
  \begin{eqnarray}
    -\Lambda_{ij}(\bq) & = & 2\bee (\gamma\rhos)^2T
    \int_{\bp\omega}
    (2p_i-q_i)(2p_j-q_j)\left[g^-(\bp,\omega)g^+g^-(\bp-\bq,\omega)
      + g^+g^-(\bp,\omega)g^+(\bp-\bq,\omega)\right] \nonumber \\
    & = & \bee T\int_{\bp}\frac{(2p_i-q_i)(2p_j-q_j)}
    {(p^2+m^2)\left[(\bp-\bq)^2+m^2\right]} \nonumber \\
    & = & \bee T\int_{\bp}\int_0^1du\frac{(2p_i-q_i)(2p_j-q_j)}
    {\left\{u(p^2+m^2)+(1-u)\left[(\bp-\bq)^2+m^2\right]\right\}^2},
  \end{eqnarray}
  with $\bee=\sum_{\alpha=1}^{N/2}e_\alpha^2$. Next, we transform to
  $\bp'=\bp-(1-u)\bq$ and expand the integral in small $\bq$, since we
  are interested in identifying its $\mathcal{O}(q^2)$ piece. As a
  result we find
  \begin{eqnarray}
    -\Lambda_{ij}(\bq\rightarrow 0) & = &
    2\bee T\delta_{ij}\int_0^1du\int_{\bp'}\left[\frac{p'^2}{(p'^2+m^2)^2} -
      \frac{2u(1-u)q^2p'^2}{(p'^2+m^2)^3}\right]
    +\bee Tq_iq_j\int_0^1du \int_{\bp'}\frac{(1-2u)^2}{(p'^2+m^2)^2} \nonumber \\
    &=&\frac{\bee T}{2\pi}\left[\log\left(\frac{\Lambda^2}{m^2}\right)-1\right]\delta_{ij}
    -\frac{\bee T}{12\pi m^2}(\delta_{ij}q^2-q_iq_j),
  \end{eqnarray}
  where $\Lambda$ is an ultra-violet cutoff on $|\bp|$.

  To calculate the transverse thermoelectric transport coefficient in
  weak magnetic fields, we need to consider diagrams with three
  legs. The direct response to a temperature gradient is
  diagramatically given in Fig. \ref{fig:triangles}a. Focusing only
  the leading terms in $Q_iq_j$, we find,
  \begin{eqnarray}
    \Delta_{\delta T}& = & \sum_\alpha\int_{\bp\omega}2\gamma
    g^-(\bp,\omega)\ea\rhos(2p_i-Q_i-q_i)g^+(\bp-\bQ-\bq,\omega)
    \gamma\ea\rhos(2p_j-Q_j-2q_j)g^+(\bp-\bq,\omega) \nonumber \\
    & & + \sum_\alpha\int_{\bp\omega}2\gamma g^-(\bp+\bq,\omega)\gamma\ea\rhos(2p_j+Q_j+2q_j)
     g^-(\bp+\bQ+\bq,\omega)\ea\rhos(2p_i+Q_i+q_i)g^+(\bp,\omega) \nonumber \\
    & = & 4\bee \int_{\bp}\frac{(2p_i-Q_i-q_i)(2p_j-Q_j-2q_j)}
    {\left[p^2+(\bp-\bQ-\bq)^2+2m^2\right]\left[p^2+(\bp-\bq)^2+2m^2\right]}
    \nonumber \\ & = & 4\bee \int_0^1du\int_{\bp}\frac{(2p_i-Q_i-q_i)(2p_j-Q_j-2q_j)}
    {\left\{u\left[2p^2+2m^2-2\bp\cdot(\bQ+\bq)+(\bQ+\bq)^2\right]+(1-u)
        \left[2p^2+2m^2-2\bp\cdot\bq+q^2\right]\right\}^2} \nonumber \\
    & = &\bee \int_0^1du\int_{\bp}\frac{(2p_i-Q_i-q_i)(2p_j-Q_j-2q_j)}
    {\left[p^2+m^2-\bp\cdot(u\bQ+\bq)+\mathcal{O}(q^2,Q^2,\bq\cdot\bQ)
    \right]^2} \nonumber \\
     & \approx & \delta_{ij}(\cdots)+q_iQ_j(\cdots) + \bee Q_iq_j\int_0^1du
     \int_{\bp'}\frac{1-u}{(p'^2+m^2)^2} \nonumber \\
    & \approx & \delta_{ij}(\cdots)+q_iQ_j(\cdots) + Q_iq_j\frac{\bee }{8\pi m^2},
  \end{eqnarray}
  where $\bp'=\bp-(u\bQ+\bq)/2$. Similarly, we need to calculate the
  response to a gradient in $\sigma$, given by the diagrams in
  Fig. \ref{fig:triangles}b. Retaining again only the leading terms in
  $Q_iq_j$, we have,
  \begin{eqnarray}
    \Delta_\sigma & = & -\sum_\alpha\int_{\bp\omega}g^-(\bp,\omega)\ea\rhos(2p_i-Q_i-q_i)g^+(\bp-\bQ-\bq
    ,\omega)\gamma\ea\rhos(2p_j-Q_j-2q_j)2\gamma Tg^+g^-(\bp-\bq,\omega)
    \nonumber \\ & & -\sum_\alpha\int_{\bp\omega}2\gamma Tg^+g^-(\bp,\omega)\ea\rhos
    (2p_i-Q_i-q_i)g^+(\bp-\bQ-\bq,\omega)\gamma\ea\rhos(2p_j-Q_j-2q_j)
    g^+(\bp-\bq,\omega) \nonumber \\ & & -\sum_\alpha\int_{\bp\omega} g^-(\bp,\omega)
    \ea\rhos(2p_i-Q_i-q_i)2\gamma Tg^+g^-(\bp-\bQ-\bq,\omega)\gamma\ea\rhos
    (2p_j-Q_j-2q_j)g^-(\bp-\bq,\omega) + {\rm c.c.} \nonumber \\
    & = & -\frac{2\bee T}{\gamma\rhos}\int_\bp\frac{(2p_i-Q_i-q_i)
      (2p_j-Q_j-2q_j)}{(p^2+m^2)\left[(\bp-\bQ-\bq)^2+m^2\right]\left[
        (\bp-\bq)^2+m^2\right]} \nonumber \\ & = &
    -\frac{4\bee T}{\gamma\rhos}\int_0^1du\int_0^{1-u}dv\int_\bp\frac{
      (2p_i-Q_i-q_i)(2p_j-Q_j-2q_j)}{\left\{p^2+m^2+u\left[(\bp-\bQ-\bq)^2-p^2
        \right]+v\left[(\bp-\bq)^2-p^2\right]\right\}^3} \nonumber \\
    & = & -\frac{4\bee T}{\gamma\rhos}\int_0^1du\int_0^{1-u}dv\int_{\bp}\frac{
      (2p_i-Q_i-q_i)(2p_j-Q_j-2q_j)}{\left\{\left[\bp-u\bQ-(u+v)\bq\right]^2
        +m^2+\mathcal{O}(q^2,Q^2,\bq\cdot\bQ)\right\}^3} \nonumber \\
    & \approx & \delta_{ij}(\cdots)+q_iQ_j(\cdots)-Q_iq_j\frac{8\bee T}
    {\gamma\rhos}\int_0^1du\int_0^{1-u}dv\,\int_{\bp'}
    \frac{(2u-1)(u+v-1)}{(p'^2+m^2)^3} \nonumber \\
    & \approx & \delta_{ij}(\cdots)+q_iQ_j(\cdots)-Q_iq_j\frac{\bee T}
    {12\pi\gamma\rhos m^4}.
  \end{eqnarray}

\end{widetext}


\begin{thebibliography}{999}
\bibitem{decouplenote} This transformation is mathematically well
  defined provided the integration contour at infinity is along the
  line ${\rm Im}(\lambda+\sigma)={\rm Re}(\lambda-\sigma)=0$.
\bibitem{FDTnote} For the model to obey the fluctuation-dissipation
  theorem in the limit $N\to\infty$, it is required that the the $\na^2$
  terms in the action vanish. The given solution ensures this to the
  desired accuracy.
\bibitem{Exnote} It is a rather difficult task to calculate the $1/N$
  corrections themselves, due to the non-trivial analytical form of
  the polarization diagrams in Figs. \ref{fig:bubbles}b,c as a
  function of frequency $\Omega$.
\end{thebibliography}
\end{document}